\documentclass[12pt]{article}

\usepackage{amssymb}
\usepackage{color}
\usepackage{epsfig,amssymb,amsfonts,amsmath,graphicx,dsfont,cite,xfrac}
\usepackage{authblk}
\usepackage{subfigure}
\definecolor{mygray}{gray}{0.5}

\usepackage{cite}
\usepackage[colorlinks=true,linkcolor=blue,citecolor=red]{hyperref}

\newcommand{\be}{\begin{equation}}
\newcommand{\ee}{\end{equation}}
\newcommand{\bea}{\begin{eqnarray}}
\newcommand{\eea}{\end{eqnarray}}

\title{Fractional Driven Damped Oscillator}

\author[${}$]{Fernando Olivar-Romero}
\author[${}$]{Oscar Rosas-Ortiz}

\affil[${}$]{\footnotesize Physics Department, Cinvestav, AP 14-740, 07000
M\'exico City, Mexico}

\date{}
\begin{document}

\maketitle

\begin{abstract}
The resonances associated with a fractional damped oscillator which is driven by an oscillatory external force are studied. It is shown that such resonances can be manipulated by tuning up either the coefficient of the fractional damping or the order of the corresponding  fractional derivatives.
\end{abstract}


\section{Introduction}

The simplest oscillating system (a {\em harmonic oscillator}) can be modeled by a mass at the end of a spring which slides back and forth without friction. The motion is characterized by the natural frequency of oscillation $\omega_0$ and the total stored energy $E$ (which is a constant of motion and defines the amplitude of oscillation) \cite{Fre71}. Actual oscillating systems present some loss of energy due to friction forces so that the amplitude of their oscillations is a decreasing function of time. However, the oscillations can be driven to avoid their damping down by the action of a repetitive force $F(t)$ on the system. Such a system is called {\em driven damped oscillator} \cite{Tay05}. An special excitation of the system arises when the frequency of the applied force matches the natural frequency of the oscillator since the spectral energy distribution takes its maximum value. The phenomenon, known as {\em resonance}, is a subject of study in classical mechanics, electromagnetism, optics, acoustics and quantum mechanics, among other physical theories \cite{Ros08}. 

The present work is addressed to the study of the driven damped oscillator in the context of fractional calculus \cite{Mil93,Pod99,Hil00}. That is, the second-order differential equation associated with the Newtonian law of motion for a damped oscillator that is driven by an external force will be substituted by a fractional differential equation of order $2\alpha$, with $0 < \alpha \leq 1$. Special emphasis will be placed on the resonance phenomenon.

\section{Fractional Oscillator with fractional damping}

Given an oscillator of natural frequency  $\omega_{0}$, the general expression for the displacement of the mass can be expressed as the integral equation \cite{Nar01,Oli14}
\begin{equation} 
x(t)=x_0 + I_t \left(\dot{x_{0}} \right) + \omega^2_0 I_t \left[ I_t \left( x(t) \right) \right],
\label{eq1}
\end{equation}
where $x_0$ and $\dot{x_{0}}$ are constants of integration, and 
\begin{equation}
I_t (x(t)) := \int x(t) dt
\end{equation}
represents the Riemman time-integration of $x(t)$. The fractional generalization of (\ref{eq1}) is performed in two steps. First we replace $I_t$ with the Riemman-Liouville fractional time-integral operator $I^{\alpha}$ \cite{Hil00,Pod99,Mil93}, and $\omega_{0}$ with $\omega_{0}^{\alpha}$. The latter for consistency of units. Then we have
\begin{equation} 
x(t)=x_{0} + I^{\alpha} \left( \dot{x_{0}} \right) + \omega^{2\alpha}_0 I^{\alpha} \left[ I^{\alpha} \left( x(t)  \right) \right], \quad 
0<\alpha \leq1.
\label{newton}
\end{equation}
Now, a fractional differential form of (\ref{newton}) can be obtained by applying twice the time-fractional derivative operator of Caputo
\begin{equation} 
D^{\alpha}D^{\alpha}x(t) +\omega^{2\alpha}x(t)=0
\label{caputo}
\end{equation}
(for details about the operator $D^{\alpha}$ see, e.g., \cite{Pod99}). Let us introduce a `fractional damping' which is proportional to the fractional time-derivative of the position $D^{\alpha} x(t)$. That is
\begin{equation} 
D^{\alpha}D^{\alpha}x(t) + 2\beta^{\alpha}D^{\alpha}x(t)+\omega^{2\alpha}x(t)=0. 
\label{caputo2}
\end{equation}
One can show that the solution of this last equation is of the form
\begin{equation} 
\begin{aligned}
x(t) = &  \frac{x_{0} t^{-\alpha} }{\sqrt{\beta^{2\alpha}-\omega_{0}^{2\alpha}}} \left[  E_{\alpha,1-\alpha}  \left( -\Omega_- t^{\alpha} \right) -  E_{\alpha,1-\alpha} \left( - \Omega_+ t^{\alpha}  \right) \right] \\ 
  & \qquad \qquad \qquad +  \frac{2\beta^{\alpha} x_{0}+x_{0}^{\left(\alpha \right)}}{\sqrt{\beta^{2\alpha}-\omega_{0}^{2\alpha}}}\left[ 
E_{\alpha,1}  \left( -\Omega_- t^{\alpha} \right) -  E_{\alpha,1} \left( - \Omega_+ t^{\alpha}  \right) \right], 
\end{aligned}
\end{equation} 
where 
\begin{equation} 
E_{\alpha,\beta}(z)=\sum_{k=0}^{\infty}\frac{z^{k}}{\Gamma ( \alpha k + \beta)}, \quad \mbox{Re}(\alpha) >0 , \quad \mbox{Re} (\beta) >0, \quad  z \in \mathbb{C},
\end{equation}
is the Mittag-Leffler function \cite{Erd55}, and 
\[
\Omega_{\pm} = \beta^{\alpha} \pm \sqrt{\beta^{2\alpha}-\omega_{0}^{2\alpha}}.
\]

\section{Driven fractional oscillator with fractional damping} 

Let us add a driving force at the right hand side of Eq.~(\ref{caputo2}), we have
\begin{equation} 
D^{\alpha}D^{\alpha}x(t) + \beta^{\alpha}D^{\alpha}x(t)+\omega^{2\alpha}x(t)=f_0 \cos\left(\omega t + \phi \right),
\label{caputo3}
\end{equation}
where $f_0$ is a real constant. Applying the Laplace transform $L$ and solving for $X(s)=L[x(t)]$ we arrive at the expression
\begin{equation} 
X(s)=f_{0}\left[ \frac{s \cos \phi-\omega \sin\phi}{\left(s^{2}+\omega^{2}\right)\left(s^{2\alpha}+2\beta^{\alpha}s^{\alpha}+\omega_{0}^{2\alpha}\right)} \right] + \frac{x_{0}s^{2\alpha-1}+\left(x_{0}+2\beta^{\alpha}x_{0}^{(\alpha)}\right)s^{\alpha-1}}{s^{2\alpha}+2\beta^{\alpha}s^{\alpha}+\omega_{0}^{2\alpha}}. 
\label{caputo4}
\end{equation}
Making $f_0=0$ we see that the second term in (\ref{caputo4}) corresponds to {\em transient} oscillations because there is no force present which can ensure their predominance; the corresponding inverse Laplace transform has been evaluated in the previous section. In turn, for the inverse Laplace transform of the firs term one has
\begin{equation} 
x(t)=\frac{f_{0}}{2\pi i}\lim_{T \to \infty}\int_{g-iT}^{g+iT}e^{st}\left[ \frac{s \cos\phi-\omega \sin\phi}{\left(s^{2}+\omega^{2}\right)\left(s^{2\alpha}+2\beta^{\alpha}s^{\alpha}+\omega_{0}^{2\alpha}\right)} \right] ds.
\label{integral}
\end{equation}
The integrand contains a branch point at $s=0$ and simple poles at $s=\pm i\omega$, and $s= \left( \Omega_{\pm} e^{\pm i\pi} \right)^{1/\alpha}$. Following \cite{Nar02} we find that $x(t)$ is given as the sum of three contributions: $x_{1}(t)$, $x_{2}(t)$ and $x_{3}(t)$. The first one results from the calculation of (\ref{integral}) along the Hankel-Bromwich path shown in Fig.~1 of Ref.~\cite{Nar02}, we obtain
{\footnotesize 
\begin{equation}  
x_{1}(t)=\frac{f_{0}}{\pi} \int_{0}^{\infty} \frac{e^{-rt}\left[rcos\phi+\omega sin\phi\right] \left[ r^{2\alpha} \sin(2\pi \alpha)+2\beta^{\alpha}r^{\alpha} \sin(\pi \alpha) \right] }
{(r^{2}+\omega^{2}) \left[ r^{4\alpha}+4\beta^{2\alpha}r^{2\alpha}+\omega^{4\alpha}+4\beta^{\alpha}r^{3\alpha} \cos(\alpha \pi)+4\beta^{\alpha}r^{\alpha}\omega_{0}^{2\alpha}+2r^{2\alpha}\omega_{0}^{2\alpha} \cos(2\alpha \pi) \right] } \, dr.
\end{equation} 
}\noindent
The latter expression vanishes as $t \rightarrow \infty$. On the other hand, the sum of residues associated with the poles $s= \left( \Omega_{\pm} e^{\pm i\pi} \right)^{1/\alpha}$ gives
{\footnotesize
\begin{multline} 
x_{2}(t)= \frac{2f_{0}e^{t\gamma_{+} \cos(\pi/\alpha)}}{\gamma_{+}^{\alpha-1
}{(\gamma_{-}^{\alpha}-\gamma_{+}^{\alpha})}}\left[\frac{\omega^{2}\gamma_{+} \cos\phi \cos\left(t\gamma_{+} \sin(\pi/\alpha)-\frac{\pi}{\alpha}(\alpha-2)\right)+\gamma_{+}^{3} \cos\phi \cos\left(t\gamma_{+} \sin(\pi/\alpha)-\pi \right)}{\omega^{4}+\gamma_{+}^{4}+2\omega^{2}\gamma_{+}^{2} \cos\left(2\pi/\alpha\right)}\right] \\
+ \frac{2f_{0}e^{t\gamma_{-} \cos(\pi/\alpha)}}{\gamma_{-}^{\alpha-1
}{(\gamma_{+}^{\alpha}-\gamma_{-}^{\alpha})}}\left[\frac{\omega^{2}\gamma_{-} \cos\phi \cos\left(t\gamma_{-} \sin(\pi/\alpha)-\frac{\pi}{\alpha}(\alpha-2)\right)+\gamma_{-}^{3} \cos\phi \cos\left(t\gamma_{-} \sin(\pi/\alpha)-\pi \right)}{\omega^{4}+\gamma_{-}^{4}+2\omega^{2}\gamma_{-}^{2} \cos\left(2\pi/\alpha\right)}\right] \\
+\frac{2f_{0}e^{t\gamma_{+} \cos(\pi/\alpha)}}{\gamma_{+}^{\alpha-1
}{(\gamma_{-}^{\alpha}-\gamma_{+}^{\alpha})}}\left[\frac{\omega^{3} \sin\phi \cos\left(t\gamma_{+}\sin(\pi/\alpha)-\frac{\pi}{\alpha}(\alpha-1)\right)+\omega^2\gamma_{+}^{2} \sin\phi \cos\left(t\gamma_{+} \sin(\pi/\gamma)-\frac{\pi}{\alpha}(\alpha + 1)\right)}{\omega^{4}+\gamma_{+}^{4}+2\omega^{2}\gamma_{+}^{2} \cos\left(2\pi/\alpha\right)}\right] \\
+\frac{2f_{0}e^{t\gamma_{-} \cos(\pi/\alpha)}}{\gamma_{-}^{\alpha-1
}{(\gamma_{+}^{\alpha}-\gamma_{-}^{\alpha})}}\left[\frac{\omega^{3} \sin\phi \cos\left(t\gamma_{-}\sin(\pi/\alpha)-\frac{\pi}{\alpha}(\alpha-1)\right)+\omega^2\gamma_{-}^{2} \sin\phi \cos\left(t\gamma_{-} \sin(\pi/\gamma)-\frac{\pi}{\alpha}(\alpha + 1)\right)}{\omega^{4}+\gamma_{-}^{4}+2\omega^{2}\gamma_{-}^{2} \cos\left(2\pi/\alpha\right)}\right],
\label{trans}
 \end{multline} 
}\noindent where $\gamma_{\pm} = \Omega_{\pm}^{1/\alpha}$. The term $x_2(t)$ is parameterized by the order of the Caputo operator $D^{\alpha}$, the coefficient $\beta$ of the fractional damping, and the natural frequency $\omega_0$; the appropriate combination of these three parameters produces $x_2 \rightarrow 0$ as $t \rightarrow \infty$. Further details will be reported elsewhere. On the other hand, the term associated with the poles $s=\pm i\omega$  is of the form
\begin{equation} 
x_{3}(t)=A \cos(\omega t + \delta),
\end{equation}
where the amplitude and phase are respectively given by

{\footnotesize
\begin{equation} 
A=f_{0} \frac{ 
\sqrt{
\omega^{4\alpha} + \omega_{0}^{4\alpha} + 2 \omega^{2\alpha}\omega_{0}^{2\alpha}\cos \left( \pi \alpha - 2\phi \right) + 4 \omega^{\alpha} \beta^{\alpha} \left[
\omega^{\alpha} \beta^{\alpha}  + \left( \omega^{2\alpha} +\omega_{0}^{2\alpha} \right)
\cos\left(\frac{\pi \alpha}{2} -2\phi \right)
\right]} }{
\omega^{4\alpha} + \omega_{0}^{4\alpha}+ 2\omega^{2\alpha}\omega_{0}^{2\alpha} \cos\left( \pi \alpha \right) + 4\omega^{\alpha}\beta^{\alpha} \left[
\omega^{\alpha}\beta^{\alpha}  + \left(  \omega^{2\alpha} + \omega_0^{2\alpha} \right)
\cos \left( \frac{\pi \alpha}{2}\right) \right]
}, 
\label{amplitude}
\end{equation} 
}\noindent and
\begin{equation} 
\delta= \arctan\left[ 
-\frac{\omega^{2\alpha} \sin\left( \pi \alpha - \phi \right) + \omega_{0}^{2} \sin\left( \phi \right) + 2\beta^{\alpha}\omega^{\alpha} \sin\left(\frac{\pi \alpha}{2}\right)}{\omega^{2\alpha} \cos\left( \pi \alpha - \phi \right) + \omega_{0}^{2} \cos\left( \phi \right) + 2\beta^{\alpha}\omega^{\alpha}\cos\left(\frac{\pi \alpha}{2}\right)} 
\right]. 
\end{equation} 
As in the conventional case, the amplitude $A$ of the oscillations dictated by $x_3(t)$ is proportional to the amplitude $f_0$ of the driving force. At zero frequency $\omega$ (i.e., for a constant driving force), the quotient $\Lambda=A/f_0$ becomes $\omega_0^{-2\alpha}$ which, in turn, reproduces the (low frequencies) Newtonian result for $\alpha=1$. At very high frequencies we find $\Lambda \approx \omega^{-2\alpha}$, so that the external driving force is dominant. On the other hand, at $\omega =\omega_0$ with $\beta$ and $\phi$ fixed, we find that $\Lambda$ is as larger as $\alpha$ approaches the value $\alpha=1$ and becomes smaller for $\alpha \in (0, 1/2)$. That is, given the fractional damping parameter $\beta$, the fractional system behaves as an underdamped oscillator for $\alpha \rightarrow 1$, and as an overdamped one if $\alpha$ approximates $1/2$ from above.

\begin{figure}[htb]
\centering\includegraphics[scale=0.5]{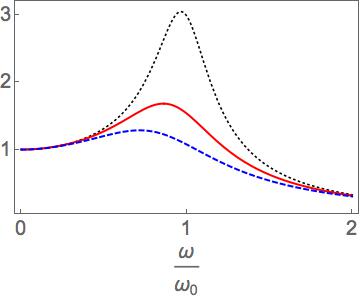}

\caption{\footnotesize (Color online) The quotient $\Lambda=A/f_0$ of the amplitude of oscillation $A$ defined in (\ref{amplitude}) and the amplitude $f_0$ of the driving force introduced in (\ref{caputo3}) with $\phi=0$, $\omega_{0}=1$, $\beta= 0.1 \omega_{0}$, for  $\alpha=0.99$ (dotted-black), $\alpha=0.95$ (solid-red), and $\alpha=0.90$ (dashed-blue).
}
\label{fig1}
\end{figure}

The behavior of $\Lambda$ is depicted in Fig.~\ref{fig1} as a function of the frequency $\omega$ with $\phi$, $\omega_0$ and $\beta$ fixed, and for different values of $\alpha$. For $\alpha = 1$ (i.e., for the Newtonian case) we find that $\Lambda$ reaches its maximum value when the driving force oscillates at the natural frequency $\omega_0$, as expected. Such a large response of the system to the driving force is the fingerprint of a resonance. Notice however that the maximum decreases and shifts to the left as $\alpha$ decreases. That is, for $\alpha \lesssim 1$ the resonance occurs at a frequency $\omega$ which is lower than $\omega_0$. The latter means that  the resonances can be controlled by fixing the fractional-damping parameter $\beta$ and tuning up the order of the fractional derivative $D^{\alpha}$. The same holds if one fixes the value of $\alpha$ and adjust the fractional-damping parameter~$\beta$.

\section{Conclusions}

Using fractional calculus one finds that the classical harmonic oscillator is affected by an `intrinsic damping' \cite{Nar01,Oli14}, such a damping is also present in the quantum-fractional case \cite{Oli16}. The response of the classical fractional oscillator to the presence of a driving force has been already studied in e.g.~\cite{Nar02}. In this paper we have presented some preliminary results of our study on a driven fractional oscillator which is affected by a fractional damping of the form $ \beta^{\alpha} D^{\alpha}x(t)$, with $D^{\alpha}$ the Caputo time-derivative operator and $0 < \alpha \leq 1$. In particular, we have shown that the resonance phenomenon can be controlled by tuning up either the coefficient $\beta$ of the fractional-damping or the order $\alpha$ of the Caputo operator. Further results will be reported elsewhere.

\section{Acknowledgments}

F.O.R. acknowledges the funding received through a CONACyT scholarship.



\begin{thebibliography}{99}


\bibitem{Fre71} 
French A P, \textit{Vibrations and waves}, W W Norton, New York, 1971

\bibitem{Tay05} 
Taylor J R, \textit{Classical Mechanics} University Science Books 2005

\bibitem {Ros08} 
Rosas-Ortiz O, Fern\'andez Garc\'ia N, Cruz~y~Cruz S, {\em  AIP Conf. Proc.} \textbf{1077} (2008) 31

\bibitem{Mil93} 
Miller K S and Ross B, \textit{An introduction to the fractional calculus and fractional differential equations}, John Wiley, New York, 1993

\bibitem{Pod99} 
Podlubny I, \textit {Fractional Differential Equations}, Academic Press, New York, 1999

\bibitem{Hil00} 
Hilfer R, \textit{Applications of Fractional Calculus in Physics}, World Scientific, New York, 2000

\bibitem{Erd55}
Erd\' elyi A, {\em Higher Transcendental Functions}, Vol. III, McGraw-Hill, New York, 1955

\bibitem{Nar01} 
Narahari A B N, Hanneken J W, Enck T and Clarke T, {\em  Physica  A} {\bf  297} (2001) 361

\bibitem{Oli14} 
Olivar-Romero F, \textit{A first approach to the fractional quantum mechanics}, M.Sc. Thesis (in Spanish), Physics Department, Cinvestav, M\'exico City, 2014

\bibitem{Nar02} 
Narahari A B N, Hanneken J W and Clarke T, {\em Physica A} {\bf 309} (2002) 275

\bibitem{Oli16}
Olivar-Romero F and Rosas-Ortiz O, {\em J. Phys. Conf. Ser.} {\bf 698} (2016) 012025


\end{thebibliography}
\end{document}